\documentstyle [emulateapj]{article}

 1

\def\etal{{et al.~}}

\def\cm{{\rm cm}}

\def\nH{{\rm H}}
\def\nHII{{\rm HII}}

\def\HII{\hbox{H~$\scriptstyle\rm II\ $}}

\def\spose#1{\hbox to 0pt{#1\hss}}
\def\lta{\mathrel{\spose{\lower 3pt\hbox{$\mathchar"218$}}
     \raise 2.0pt\hbox{$\mathchar"13C$}}}
\def\gta{\mathrel{\spose{\lower 3pt\hbox{$\mathchar"218$}}
     \raise 2.0pt\hbox{$\mathchar"13E$}}}

\def\gs{\mathrel{\raise1.16pt\hbox{$>$}\kern-7.0pt
\lower3.06pt\hbox{{$\scriptstyle \sim$}}}}
\def\ls{\mathrel{\raise1.16pt\hbox{$<$}\kern-7.0pt
\lower3.06pt\hbox{{$\scriptstyle \sim$}}}}
\def\gtsima{$\; \buildrel > \over \sim \;$}
\def\ltsima{$\; \buildrel < \over \sim \;$}
\def\prosima{$\; \buildrel \propto \over \sim \;$}
\def\gsim{\lower.7ex\hbox{\gtsima}}
\def\lsim{\lower.7ex\hbox{\ltsima}}
\def\simgt{\lower.7ex\hbox{\gtsima}}
\def\simlt{\lower.7ex\hbox{\ltsima}}
\def\simpr{\lower.7ex\hbox{\prosima}}

\def\pp{\noindent\parshape 2 0truecm 17truecm 2truecm 15truecm}
\def\rf#1;#2;#3;#4 {\par\pp#1, #2, #3, #4. \par}

\def\pr{\ref@jnl{Phys.Rev}}     
\def \ptp #1 #2 {{ Progr. Theoret. Phys.\/} {#1},  {#2}}

\def\href#1;#2 {{\bf #1} : {\em #2}}

\def\beq#1{\begin{equation}\label{#1}}
\def\eeq{\end{equation}}
\def\beqa#1{\begin{eqnarray}\label{#1}}
\def\eeqa{\end{eqnarray}}
\def\eq#1{equation~(\ref{#1})}

\def\tento#1{\times 10^{#1}}

\def\Ms{\ M_{\odot}}

\def\s{{\rm \ s}}

\def\cm{{\rm \ cm}}

\def\Mpc{{\rm \ Mpc}}
\def\kpc{{\rm \ kpc}}

\def\H2p{H$_2^+$ }

\def\mH2p{H_2^+}

\makeatletter

\newenvironment{figurehere}
  {\def\@captype{figure}}
  {}
\makeatother

\begin{document}
\thispagestyle{empty}

\title {Photon Conserving Radiative Transfer around Point Sources in multi--dimensional Numerical Cosmology} 
\author {Tom Abel\altaffilmark{1,2}, Michael
  L. Norman\altaffilmark{1,3}, and Piero Madau\altaffilmark{4,5}}
\received{??}
\accepted{??}
\altaffiltext{1}{Laboratory for Computational Astrophysics, NCSA,
          University of Illinois at Urbana/Champaign, 405 N. Mathews
          Ave., Urbana, IL 61801.}  
\altaffiltext{2}{Max-Planck-Institut f\"ur Astrophysik,
          Karl-Schwarzschild-Strasse 1, 85748 Garching, Germany}
\altaffiltext{3}{Astronomy Department, University of Illinois at
    Urbana--Champaign, Urbana.}
\altaffiltext{4}{Space Telescope Science Institute, 3700 San Martin Drive,
Baltimore MD 21218.}  
\altaffiltext{5}{Institute of Astronomy, Madingley Road, Cambridge CB3 0HA, 
UK.} 
\begin{abstract}
  Many questions in physical cosmology regarding the thermal and ionization 
  history
  of the intergalactic medium are now successfully studied with the
  help of cosmological hydrodynamical simulations. Here we present a
  numerical method that solves the radiative transfer around point
  sources within a three dimensional cartesian grid. The method is
  energy conserving independently of resolution: this ensures the correct
  propagation speeds of ionization fronts.  We describe the details of
  the algorithm, and compute as first numerical application the ionized 
  region surrounding a mini-quasar in a cosmological density field 
  at $z=7$.

\end{abstract}
\setcounter{page}{1}
\section{Introduction}

The incorporation of radiation processes and radiative transfer into
cosmological hydrodynamical simulations is essential for modeling the
structure and evolution of the Ly $\alpha$ forest (Cen \etal~1994;
Zhang, Anninos \& Norman~1995; Hernquist \etal~1996; Haehnelt \&
Steinmetz~1997), interpreting the observations of helium and metal line 
absorbers at high redshift (Reimers \etal~1997; Hellsten \etal~1997; 
Songaila~1998), and for simulating the early 
reheating and reionization of the intergalactic medium (IGM) (Ostriker 
\& Gnedin~1996; Gnedin \& Ostriker~1997). In the last case the large 
computational effort is justified by the hope that the study of the 
transition from a neutral IGM to one that is almost 
fully ionized could provide some hints on the first generation of stars 
and quasars in the universe.    
Until now, radiative transfer effects have either been ignored in such
simulations, or treated as a self-shielding correction to an optically
thin approximation (Katz \etal~1996; Gnedin \& Ostriker~1997).
Typically, radiation fields are treated as isotropic backgrounds
$J_{\nu}(z)$ which are either specified as an external function
computed by other means (e.g., Haardt \& Madau~1996), or are computed
by averaging over all sources within the computational volume (Cen
1994; Gnedin \& Ostriker~1997). Only approximate treatments exist of the
reprocessing of radiation via absorptions internal to the sources as well as 
by the IGM, an effect that can significantly influence the spectrum of 
the metagalactic flux (Haardt \& Madau~1996). Moreover, at $z
\gta$ 3 (5) the IGM itself is opaque in the helium (hydrogen) Lyman
continuum, and radiation backgrounds become increasingly inhomogeneous
and anisotropic (Reimers \etal~1997; Madau, Haardt \& Rees~1998).
Recent analytic work has attempted to forge a closer connection
between sources, transport, and sinks of cosmic radiation, albeit
in a spatial-- and angle--averaged way (Haiman \& Loeb,~1998a, b;
Madau, Haardt \& Rees~1998).

In this {\sl Letter\,} we describe an approach to cosmological radiative
transfer which relaxes these assumptions, and is appropriate on scales
comparable to the separation of individual sources of radiation. Our
method has the property that energy is conserved independently of
numerical resolution, ensuring, for example, that ionization fronts
propagate at the correct speeds. The radiation driven front surrounding
a mini-quasar in a cosmological density field at $z=7$ is computed 
as  a first application of the technique.

\section{Cosmological Radiative Transfer}

The equation of cosmological radiative transfer in comoving coordinates
(cosmological, not fluid) is (Norman, Paschos \& Abel~1998):
\begin{equation}
\frac{1}{c} \frac{\partial I_{\nu}}{\partial t} + \frac{\hat{n} \cdot
\nabla I_{\nu}}{\bar{a}} - \frac{H(t)}{c} (\nu \frac{\partial I_{\nu}}
{\partial \nu} - 3 I_{\nu}) = \eta_{\nu} - \chi_{\nu} I_{\nu}
\label{equ:crt}
\end{equation}
where $I_{\nu} \equiv I(t, \vec{x}, \hat{n}, \nu)$ is the
monochromatic specific intensity of the radiation field, $\hat{n}$ is
a unit vector along the direction of propagation of the ray; $H(t)
\equiv \dot{a}/a$ is the (time-dependent) Hubble constant, and
$\bar{a} \equiv \frac{1+z_{em}}{1+z}$ is the ratio of cosmic scale
factors between photon emission at frequency $\nu$ and the present
time t.  Here $\eta_{\nu}$, $\chi_{\nu}$, and $c$ denote the
emission coefficient, the absorption coefficient, and the speed of light,
respectively.  Equation~\ref{equ:crt}
will be recognized as the standard equation of radiative transfer with
two modifications: the denominator $\bar{a}$ in the second term, which
accounts for the changes in path length along the ray due to cosmic
expansion, and the third term, which accounts for cosmological
redshift and dilution.  
In principle, one could solve equation (\ref{equ:crt}) directly 
for the intensity
at every point in $(t, \vec{x}, \hat{n}, \nu)$ space 
given $\eta$ and $\chi$. However the high dimensionality
of the problem not to mention the high spatial and angular resolution needed
in cosmological simulations make this approach impractical. Therefore
we proceed through a sequence of well-motivated approximations which
reduce the complexity to a tractable level.

Comparing the third with the second term in equation (\ref{equ:crt}), 
the classical transfer equation (Kirchhoff
1860):
\begin{equation}
\frac{1}{c} \frac{\partial I_{\nu}}{\partial t} + \hat{n} \cdot
\nabla I_{\nu} = \eta_{\nu} - \chi_{\nu} I_{\nu}
\label{equ:trt}
\end{equation}
\noindent is recovered if the scale of interest, $L$, is much smaller
than the horizon, $c/H(t)$. (Note that this is only true if 
$|\nu \frac{\partial I_{\nu}}{\partial \nu}| \leq I_{\nu}$; i.e., 
for continuum radiation. However, we can nonetheless still use 
eq. (\ref{equ:trt}) for line transfer provided we Doppler shift 
the absorption cross section in $\chi_{\nu}$ due to Hubble expansion.)
In the case of constant absorption and
emission coefficients the time derivative can also be dropped and one
is left with the static transfer equation
\begin{equation}
  \label{eq:srt}
  \hat{n} \cdot\nabla I_{\nu} = \eta_{\nu} - \chi_{\nu} I_{\nu}. 
\end{equation}
This equation is also adequate for problems in which the absorption
and emission coefficients change on timescales much longer than the light
crossing time, $L/c$. This will always be the case in the volumes
we will be able to simulate in the near future, and thus we adopt this
approximation. At small distances from the source, however, this
simplification will always break down allowing I--fronts to expand
faster than the speed of light. This can be easily dealt with as we
will show in the subsequent section.

An extensive literature exists on methods of solution for equation
(\ref{eq:srt}) (e.g., Mihalas \& Mihalas~1984) depending on the 
symmetry of the problem and the properties of $\eta_{\nu}$
and $\chi_{\nu}$. 
A direct solution of equation (\ref{eq:srt}) is impractical due
to the high dimensionality of the problem.
Our approach is approximate, and
is based on decomposing the radiation field into point source 
and diffuse components:
$I_{\nu} \equiv I^{pts}_{\nu} + I^{diff}_{\nu}$. 
The motivation for doing this is that whereas $I^{pts}_{\nu}$
is highly anisotropic, we expect $I^{diff}_{\nu}$ to be
nearly isotropic, and accordingly we can employ the best
methods taylored for each component. 

Utilizing the linearity of the radiation field, we can write:
\begin{eqnarray}
\hat{n} \cdot\nabla I^{diff}_{\nu} &=& \chi_{\nu}(S_{\nu} - I^{diff}_{\nu}) \\
\label{eq:diff}
\hat{n} \cdot\nabla I^{pts}_{\nu} &=& - \chi_{\nu} I^{pts}_{\nu},
\label{eq:ray}
\end{eqnarray}
\noindent
where diffuse emission (e.g., due to recombination
radiation) appears only in the transport equation for the diffuse radiation
field via the source function $S_{\nu}\equiv \eta_{\nu}/\chi_{\nu}$. 
This has the advantage that equation (\ref{eq:ray}) can now be solved 
ray by ray in a completely decoupled fashion.
For $\chi_{\nu}(x) = const$, equation (\ref{eq:ray}) 
has the simple analytic solution
\begin{eqnarray}
  \label{eq:rtsolution}
  I_{\nu}(x_1) = I_{\nu}(x_0) e^{-\tau} = I_{\nu}(x_0) \exp( -  \chi_\nu
  (x_1 - x_0)). 
\end{eqnarray}
In the next section we describe an efficient implementation for
solving equation (\ref{eq:ray}) on rays emanating from point sources
within a uniform Cartesian grid. Several methods of solution for
equation (\ref{eq:diff}) are discussed in Norman, Paschos \& Abel (1998).

\section{Implementation}

Figure 1 gives the flowchart of the simple algorithm; details of
the individual computational steps are given in the following sections.
For each point source in our volume, a set of radial rays quasi-uniformly
distributed in solid angle are constructed. 
Enough rays are used such that every cell at large distances from
the source is crossed by at least one ray {\em on average}. 
We are able to use fewer angles than one would naively assume
by Monte Carlo sampling in angle within a radiation-matter
coupling timescale.        
Each ray is discretized ``cast" into ray segments according to how
it intersects the cell boundaries. On each ray segment the
intensity is attenuated according to equation (\ref{eq:ray}) using
an absorption coefficient appropriate to that cell.  
The number of absorptions in the frequency interval $\nu, \nu + d\nu$
in each ray segment inside a cell is simply related to the
decrease in $I^{pts}_{\nu}$ along that segment.
The total number of photoionizations in the cell is the sum over
all ray segments crossing the cell.

\vspace{0.2cm}
\begin{figurehere}
\epsscale{0.2} \plotone{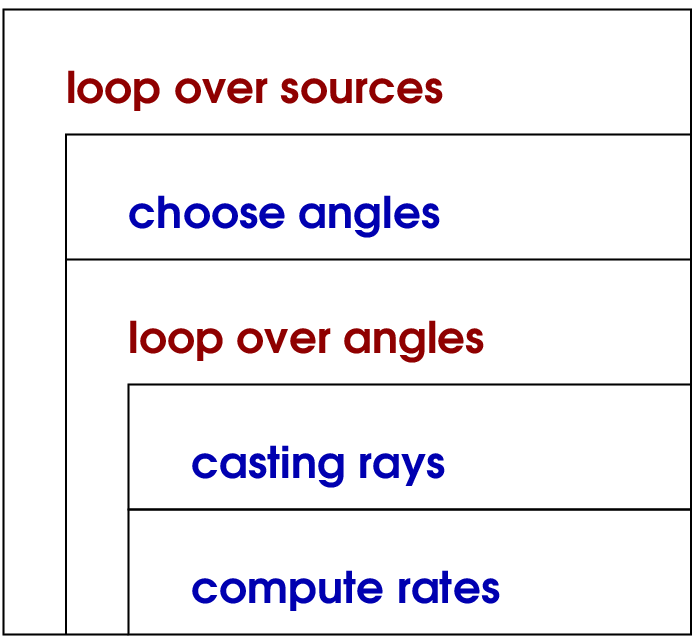} 
\caption{ \footnotesize 
  The flowchart for the radiation transfer algorithm.} 
 \label{fig:flowchart}\vspace{0.1cm}
\end{figurehere}

\subsection{Choosing Angles} 
To properly describe the I--front one needs at least $N_a= f_a \times 2\pi
r/\triangle x$ rays so that at least one ray is cast to each cell of
sidelength $\triangle x$ at the equator of a sphere with radius $r$.
We introduce the factor $f_a$ to allow us to control the number of rays. 
To adopt the minimum number of rays needed we store the
maximum radius, $r_{max}$, needed to capture the furthest point of the
I--front.

We use spherical coordinates $(r, \phi, \theta)$,
\begin{eqnarray}
  \label{eq:sphcoord}
  x &=& r \cos\phi\, \cos\theta, \ \ 0\leq\phi\leq2\pi \nonumber \\
  y &=& r \sin\phi\, \cos\theta, \ \ -\frac{\pi}{2}\leq
  \theta \leq \frac{\pi}{2} \nonumber \\
  z &=& r \sin\theta, 
\end{eqnarray}
and divide the sphere in segments of similar size such that all rays 
will have similar fluxes. 
We can now find the angles that define the rays which pierce through
segments of roughly equal area. The discrete values for $\theta$ are
given by
\begin{eqnarray}
  \label{eq:thetaj}
  \theta_j = (j-\frac{1}{2}) \frac{2\pi}{N_a} - \frac{\pi}{2},\ \
  1\leq j\leq \frac{N_a}{2}.
\end{eqnarray}
Closer to the poles fewer azimuthal angles are required,
\begin{eqnarray}
  \label{eq:Nphi}
  N_\phi^j=\max(N_a\, \cos\theta,1).
\end{eqnarray}
The azimuthal angles are now chosen to be,
\begin{eqnarray}
  \label{eq:phii}
  \phi_i=(i-\frac{1}{2}) \frac{2\pi}{N_\phi},\ \
  1\leq i\leq N_\phi.
\end{eqnarray}
One also needs to know the area of the sphere segment, $A(i,j)$
described by each ray. In units of the surface of the sphere ($4\pi
r^2$), this is
\begin{eqnarray}
  \label{eq:area}
  A(i,j)& = &\frac{|\sin\theta_2-\sin\theta_1|}{2N_\phi^j}, \nonumber \\
  \theta_1 &=& (j-1) \frac{2\pi}{N_a} - \frac{\pi}{2},\   
  \theta_2 =  j    \frac{2\pi}{N_a} - \frac{\pi}{2}.
\end{eqnarray}

\vspace{0.2cm}
\begin{figurehere}
\epsscale{0.2} \plotone{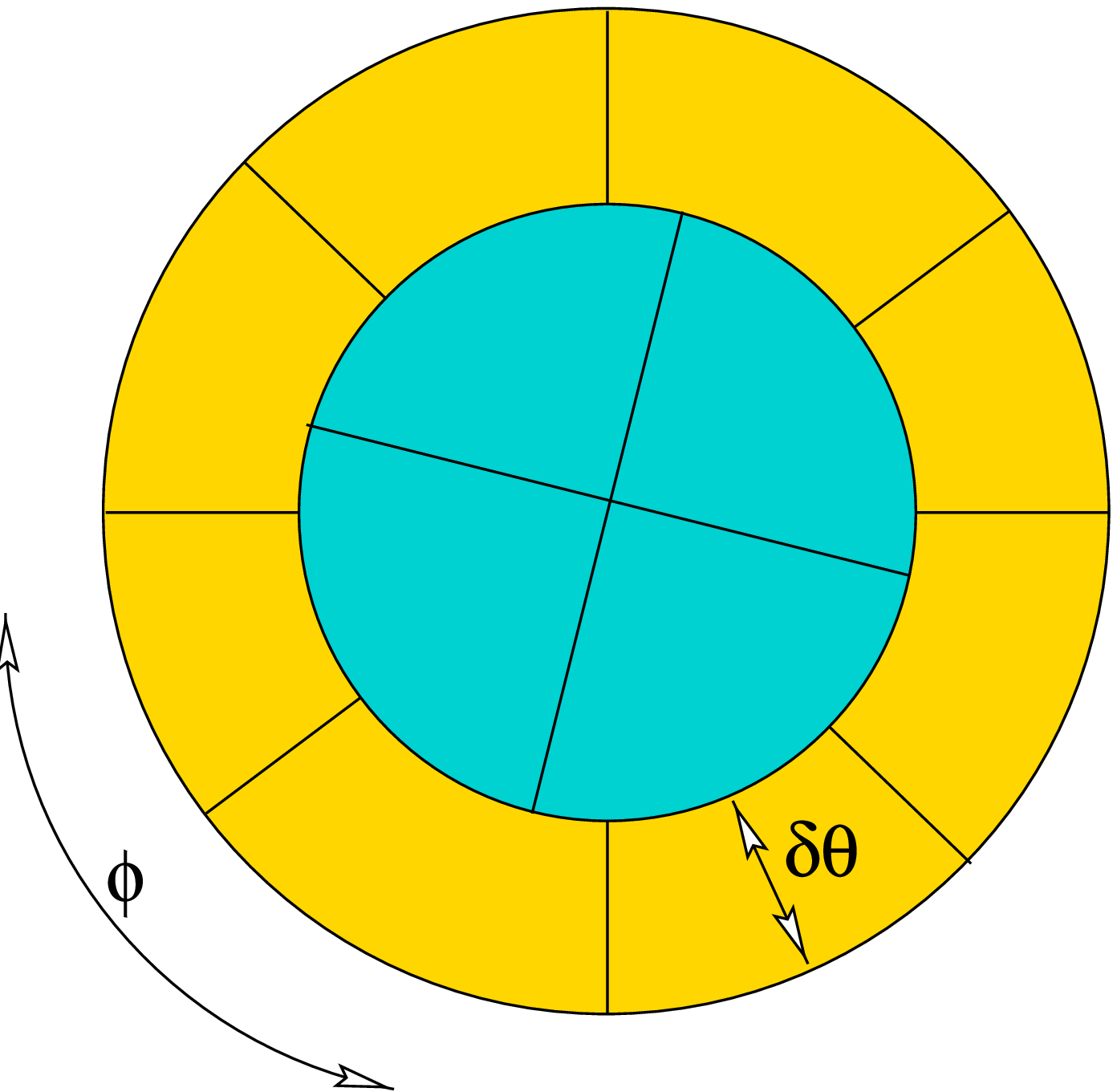} 
\caption{ \footnotesize 
  Sphere from top with the division into areas of similar
  size. The azimuthal angle $0 \leq \phi \leq 2\pi$ and $\delta
  \theta$ are also indicated. The variying $N_\phi(\theta)$ is
  evident. } 
 \label{fig:sphere}\vspace{0.1cm}
\end{figurehere}

A random rotation of the coordinate system is then introduced 
for each snapshot solution of $I_{\nu}$ to avoid
any possible pole artifacts and reduce the number of angles required.

\subsection{Casting Rays}

Ray casting is the next step. This means to compute, for a given ray
$(\theta,\phi)$, the indices $i(l), j(l), k(l)$ of the grid cells 
it traverses, as well as the lengths of the ray segments $dS(l)$
within each cell.  Here $i,j,k$ are the indices of grid cells in a
Cartesian lattice $x(i), y(j),$ and $z(k)$, and $l$ is the index of
the ray segment.  For simplicity we restrict our discussion to uniform, 
isotropic grids such that $x(i)=i\Delta, y(j)=j\Delta,
z(k)=k\Delta$ where $\Delta$ is the cell size and $0 \leq i \leq N_i$,
and similarly for j and k. 
Ray casting is a problem in computational geometry
with efficient methods discussed in books on computer graphics.
For a uniform Cartesian grid the problem is straightforward.
Here we write down a direct algorithm (i.e., requiring no costly sorts).
Consider a point
source at coordinate $x_s,y_s,z_s$ in cell $i_s,
j_s, k_s$.  We have simply $i(1)=i_s, j(1)=j_s,
k(1)=k_s$.  We cast a ray in direction $\theta,\phi$ and ask which
cell boundary it intersects first. If it intersects an x boundary
first, $i(2) = i(1) + {\rm sign}(\cos \phi), j(2)=j(1), k(2)=k(1)$,
and analogously if the y or z boundaries are crossed first. To
determine which cell face is crossed first, we compute the distances
(radii) from the source to the next x, y or z crossings.  Call these
distances $r_x, r_y$ and $r_z$.  The minimum of these three distances
determines which cell boundary is crossed first.  

This can be expressed as follows:
\noindent
Define
\begin{eqnarray}
\mu &= \cos\phi, \ \ s_{\mu} \equiv& {\rm sign}(\mu) \nonumber \\
\gamma &= \sin\phi, \ \ s_{\gamma} \equiv& {\rm sign}(\gamma)
\nonumber \\
\zeta &= \sin\theta, \ \ s_{\zeta} \equiv& {\rm sign}(\zeta) 
\end{eqnarray}
\noindent
then, for all $l \geq 2$ we have
\begin{eqnarray}
r_x(l) &=& |\Delta [i(l)+\frac{1}{2}(s_{\mu}+1)] - x_s|/\max(\epsilon, |\mu\cos\theta |) \nonumber \\
r_y(l) &=& |\Delta [j(l)+\frac{1}{2}(s_{\gamma}+1)] - y_s|/\max(\epsilon, |\gamma\cos\theta|) \nonumber \\
r_z(l) &=& |\Delta [k(l)+\frac{1}{2}(s_{\zeta}+1)] - z_s|/\max(\epsilon, |\mu\,\zeta |)
\end{eqnarray}
\begin{eqnarray}
i(l) &=& i(l-1) \nonumber \\
j(l) &=& j(l-1) \nonumber \\
k(l) &=& k(l-1) \nonumber \\
{\rm if\ } [r_x(l) \leq \min(r_y(l),r_z(l))],&&{\rm then }\nonumber \\
 S(l) &=& r_x(l) \nonumber \\
 i(l) &=& i(l-1)+s_{\mu} \nonumber \\
{\rm if\ }[r_y(l) \leq \min(r_x(l),r_z(l))],&&{\rm then }\nonumber \\
 S(l) &=& r_y(l) \nonumber \\
 j(l) &=& j(l-1)+s_{\gamma} \nonumber \\
{\rm if\ }[r_z(l) \leq \min(r_x(l),r_y(l))],&&{\rm then }\nonumber \\
 S(l) &=& r_z(l) \nonumber \\
 k(l) &=& k(l-1)+s_{\zeta}. 
\end{eqnarray}  
\noindent
Here $\epsilon$ is a small number to avoid dividing by zero.

\subsection{Computing Rates}

Although the above technique can be applied to any absorption
process, in the following we will discuss the specific case of hydrogen 
photoionization, and compute the photoionization rate coefficient and the 
associated heating term. 
Additional species can be treated in an analogous way.
Given the path length $dS(l)=S(l)-S(l-1)$, and the absorption
coefficient in cell $(i(l), j(l), k(l))$, the optical depth for
photoionization along this path is given by $\tau_\nu^l=
\chi_{\nu}(i(l), j(l), k(l)) dS(l)$.  Further, from the solution of
the static transfer equation (\ref{eq:rtsolution}), one sees that
$\delta t \times \dot {N_{\nu}}(l-1)(1-e^{-\tau_\nu^l})$ photons are absorbed
in the time interval, $\delta t$. Hence, the rate of change of the
neutral hydrogen density, $n_\nH$, due to photoionization is simply
given by,
\begin{eqnarray}
  \label{eq:rate}
  \dot{n}_\nH= \dot {N}_\nu^{l-1}\left[ 1-e^{-\tau_\nu^l}\right]/V_{\rm 
cell},
\end{eqnarray}
where $\dot {N}_\nu$ is the monochromatic number of photons emitted by 
the source per unit time, and $V_{\rm cell}$ denotes the volume of the cell. 
The heating rate is then,
\begin{eqnarray}
  \label{eq:heatingrate}
    \dot{e}= (h\nu-1\,{\rm ryd})\, \dot{N}_\nu^{l-1}\left[1-e^{-\tau_\nu^l}
\right]/V_{\rm cell}. 
\end{eqnarray}
Both of the above rates are summed over the contributions of all rays.
Employing the analytical solution for each ray segment insures that
I--fronts move at the correct speed independent of spatial resolution.

\subsection{Faster than light I--fronts}

In deriving equation (\ref{eq:srt}) we have neglected the time dependent 
term of the transfer \eq{equ:crt}. The assumption that the light
crossing time is shorter than the ionization timescale breaks down close
to the source. As a consequence the I--front expands at a speed
exceeding the speed of light. This can also be seen from the simple
jump condition in a static universe,
\begin{eqnarray}
  \label{eq:jump}
  4\pi r_I^2 n_\nH \frac{dr_I}{dt} = \dot {N}-4\pi\alpha_B\int_0^{r_I}
  n_en_p r^2 dr,
\end{eqnarray}
where $\alpha_B$ is the recombination coefficient to the excited states
of hydrogen. In this expression, $dr_I/dt$ exceeds the speed of light 
for radii less than
$r_c=(\dot {N}/4\pi cn_\nH)^{0.5}\approx 5.3\kpc ({\dot{N}_{56}/n_\nH})^{0.5}$.
To avoid this
unphysical effect we simply do not compute rates at distances further
than $r_l=c(t-t_{\rm on})$. The radius $r_l$ is also used in the
computation of the optical depths in equations (\ref{eq:heatingrate})
and (\ref{eq:rate}). Additionally, to speed up the calculation for
radii $r<r_c$ we evolve the equations on a tenth of the light
travel time across one cell. As a consequence the time evolution of
the ionized fration close to the source will not be computed
correctly. However, we do not consider this a severe limitation 
since the neutral fraction becomes practically zero in
regions close to the source within a light travel time.

\section{A Test and an Application}

In this section we first provide a test case that demonstrates the
accuracy of the method, and then, as first application, compute the 
ionized region surrounding a ``mini-quasar'' in a cosmological density field 
at $z=7$. Mini-quasars could be common at early epochs if black holes 
can form in the first $10^8\Ms$ cold dark matter (CDM) condensations
(Haiman \& Loeb~1998b).

\subsection{Spherical I--front Test}

To test our algorithm we set up a point source in a uniform medium, and
neglect the effect of radiative recombinations. The radius of the spherically
expanding I--front can be derived analytically by balancing the number of 
emitted photons with the hydrogen atoms that are present in the ionized
volume,
\begin{eqnarray}
  \label{eq:r_ana}
  R_I(\triangle t) = \left(\frac{3 \dot {N}}{4\pi n_\nH}\triangle
  t\right)^{\frac{1}{3}}. 
\end{eqnarray}
\vspace{0.2cm}
\begin{figurehere}
\epsscale{0.45} \plotone{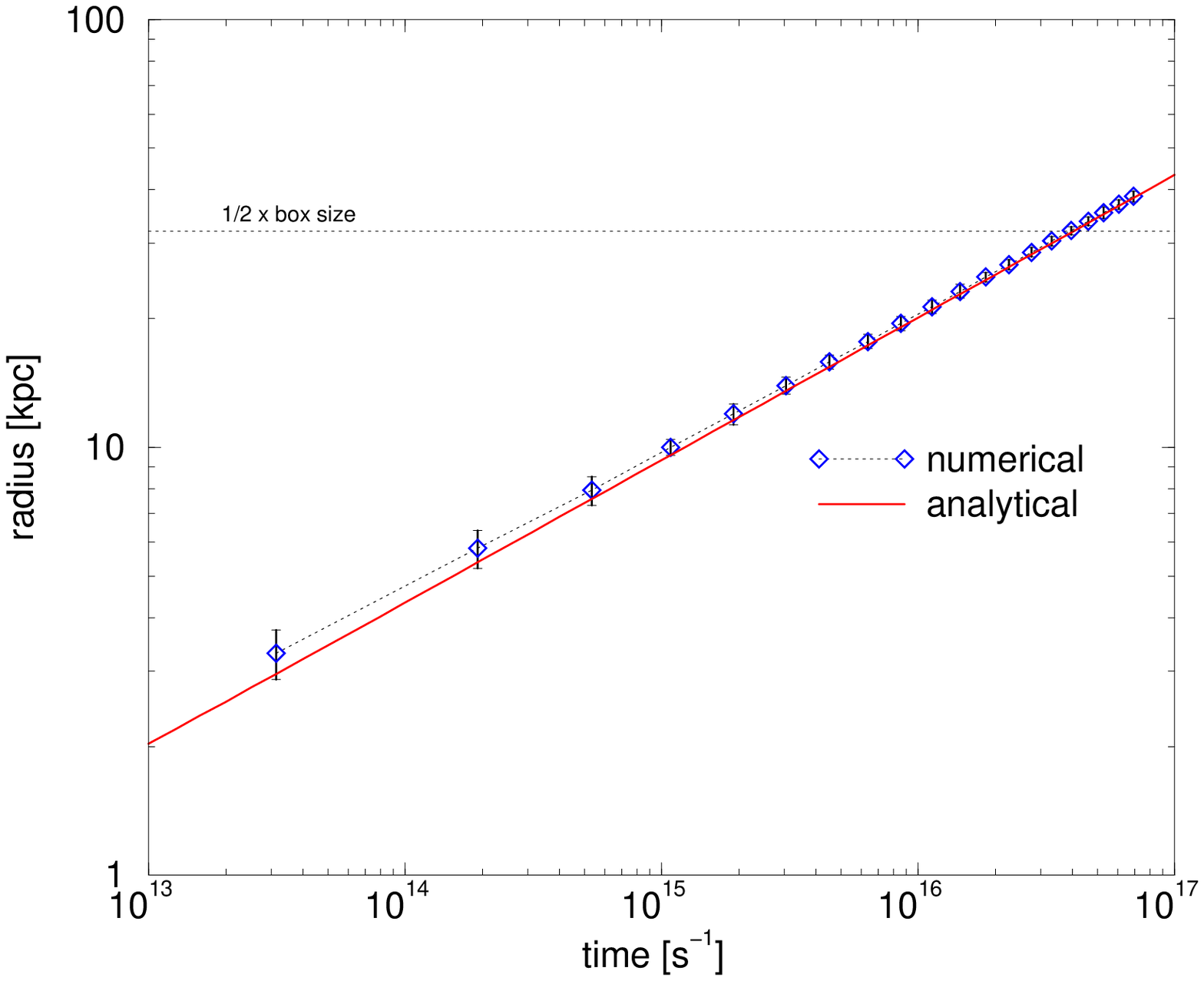} 
\caption{ \footnotesize 
  Comparison of the analytical radius of an I--front [eq. (\ref{eq:r_ana})] 
  with the numerical solution on a $64^3$ 3D
  cartesian grid. The error bars indicate the maximum deviation found
  on the spherical ionization front. The deviations are always less
  than one grid cell. The straight line indicates were part of the
  I--front leaves the simulation volume. The cell size is $1\kpc$ in 
  this test case. }
 \label{fig:radiustest}\vspace{0.1cm}
\end{figurehere}
\noindent
Figure~\ref{fig:radiustest} compares this analytical solution with
the results of our algorithm on a $64^3$ grid. The radius and the
spherical geometry are perfectly recovered within the accuracy of the
spatial resolution. The specific parameters used in this test were,
$\dot {N}=10^{51} \s^{-1}$, $n_\nH=10^{-2}\cm^{-3}$, and $\triangle
x=1\kpc$.  The algorithm has also been tested to reproduce accurately
the size of classical Str\"omgren sphere in calculations where
radiative recombinations are included.

\subsection{Cosmological Density Field}

Our first interesting application is the evolution of the ionization
zone surrounding a mini-quasar which turns on at $z=7$ in the 
inhomogeneous density field derived from cosmological simulations. 
For simplicity, we have neglected any thermal or dynamical evolution of gas, 
and included hydrogen radiative recombinations by assuming a constant Case B
recombination rate of $\alpha_B=3.4\tento{-13}$ cm$^3$ s$^{-1}$ everywhere 
on the grid. As a background medium we use the \ion{H}{1} distribution 
computed in a SCDM cosmology (with $\Omega_b=0.06$ and $h=0.5$) at
$z=6.941$ from Bryan \etal~1998. The $2.4\Mpc$ box length corresponds
to 300 proper kiloparsecs. At the densest cell, which is found in a
virialized halo of total mass $\approx 1.3\tento{11}\Ms$, we introduce a 
quasar-type source with an ionizing photon emission rate of
$\dot{N}=5\tento{53}\s^{-1}$.  The gas clumping factor is
$\langle n_\nH^2 \rangle /\langle n_\nH \rangle^2=59.2$, and is larger 
than the ionized hydrogen clumping factor, $C=\langle n_\nHII^2 \rangle
/\langle n_\nHII\rangle^2$, as clumps that are dense and thick enough to be 
self-shielded from UV radiation remain neutral and do not contribute to 
the recombination rate.

\vspace{0.2cm}
\begin{figurehere}
\epsscale{0.4} \plotone{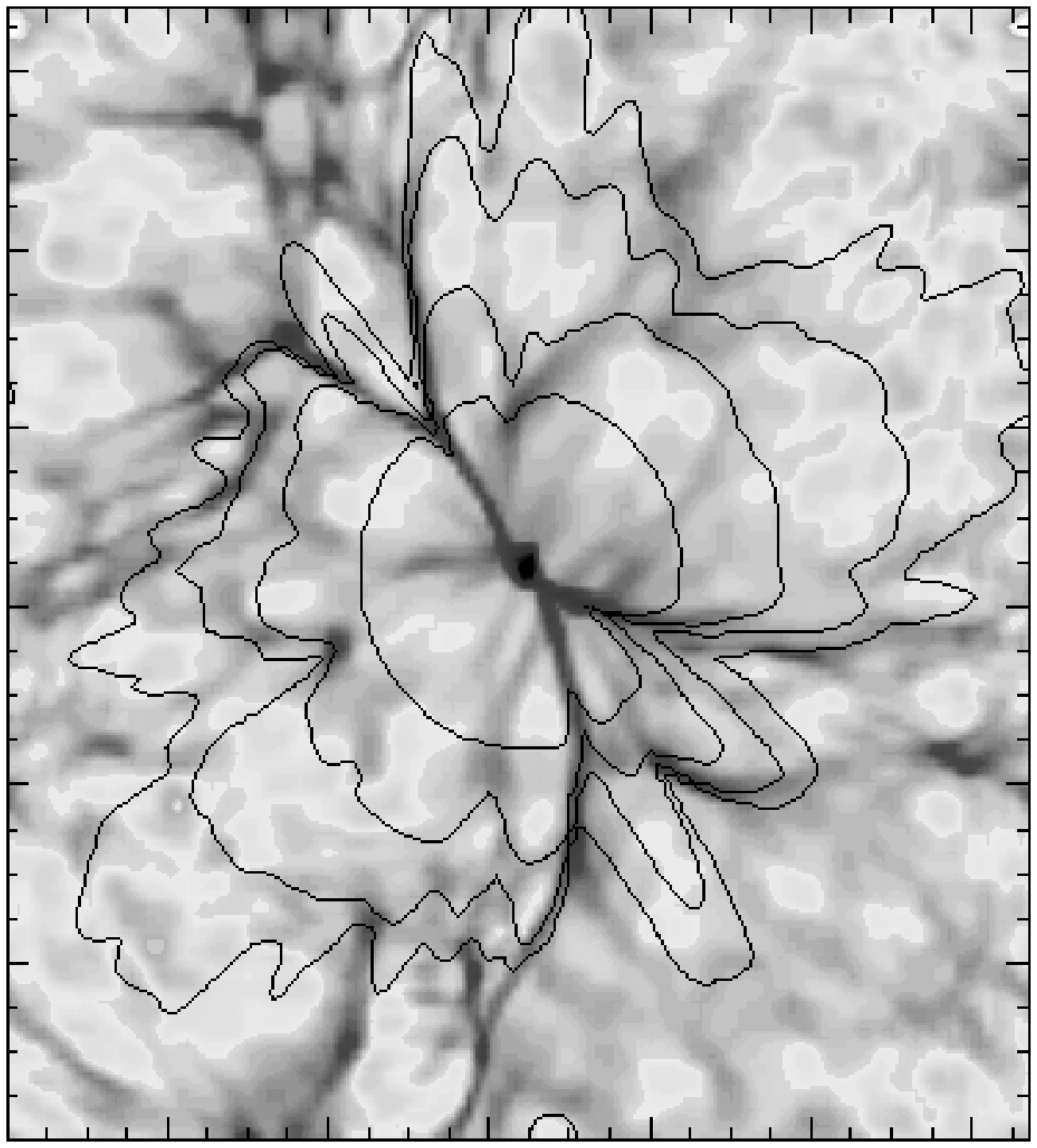} 
\caption{ \footnotesize 
  Propagation of an R-type I--front in a $128^3$ cosmological density
  field produced by a mini-quasar with $\dot {N}=5\tento{53}\s^{-1}$. 
  The solid
  contours give the position of the I--front at 0.15, 0.25, 0.38, and
  0.57\,Myr after the quasar has switched on. The underlying greyscale
  image indicates the initial \ion{H}{1} density field.}
 \label{fig:cosmotest}\vspace{0.1cm}
\end{figurehere}
\noindent
Figure~\ref{fig:cosmotest} illustrates the evolution of the I--front during
the first 0.6 Myr.
The initial spherical expansion (at the speed of light since the number of 
available photons exceeds by far the number of neutral hydrogen 
atoms and recombinations) is quickly broken as the front quickly expands 
first into the voids (increasing
their thermal pressure by many orders of magnitude by photo--heating), 
then more slowly into the denser filaments. 

In a highly inhomogeneous universe, the volume-averaged gas recombination 
timescale, 
\begin{equation}
t_{\rm rec}=(n_e\alpha_B C)^{-1}=0.1\, {\rm Gyr}
\left({1+z\over 8}\right)^{-3}
C_{20}^{-1} \label{eq:trec}
\end{equation}
(for $\Omega_bh_{50}^2=0.06$) is much shorter than the then Hubble time.
At later epochs, when the size of the \HII region is large compared to the 
scale of the clumping, the front will fill its time-varying Str\"omgren 
volume in a few recombination times just like in the static case
(Madau, Haardt, \& Rees~1998), 
\begin{equation}
V_S={\dot{N}t_{\rm rec}\over n_\nH}=0.63\, {\rm Mpc^3} \dot {N}_{53.7}
\left({1+z\over 8}\right)^{-6}
C_{20}^{-1}. 
\end{equation}
This is 20 times larger than our simulation box.

\section{Summary}

Not only for theoretical reasons, but also in light of upcoming space
missions such as MAP, PLANCK, and NGST, a detailed understanding of
the thermal history of the IGM and reionization is highly desirable.
In this paper we have described a photon conserving algorithm which
allows to simulate inhomogenous reionization self--consistently within
in a cosmological hydrodynamical simulation. The method employs an on
the spot approximation to treat the effects of diffuse emission of
ionizing photons. For scenarios in which stellar sources are
responsible for hydrogen reionization the on the spot approach is
expected to give reliable results. However, for calculations in which
diffuse radiation due to recombination of helium is also expected to
be important a separate solver can be used.  In applications where
radiative recombinations can be neglected the number of rays used can
be reduced greately due to the variable choice of coordinate system
for the selection of rays.  Recently, a differnt algorithm for 3D
radiative transfer in cosmological situation has been presented by
Razoumov and Scott (1998).  Their explicit advection scheme (at the
speed of light) seems well suited for situations in which the
I--fronts move faster than any hydrodynamical flow. Only temporary 2D
arrays are used in the method presented here requiring negligible small
additional memory for 3D cosmological hydrodynamics simulations.

\acknowledgments 
Support for this work was provided by NASA through ATP grant NAG5-4236.
T. A. also acknowledges support from NASA grant NAG5-3923. 


\vfill
\eject

\begin{references}
\reference{o} Abel, T., Anninos, P., Zhang, Y., \& Norman, M. 1997a, 
        NewA, 2, 181.

\reference{o} Abel, T., Anninos, P., Norman, M., \&  Zhang, Y. 1998, 
        \apj, in press.

\reference{o}  Bryan, G.L., Machacek, M., Anninos, P., \& Norman, M.L. 1998, 
preprint (astro-ph/9805340).

\reference{o} Cen, R., Miralda-Escud\'e, J., Ostriker, J.P., \& Rauch, M.
1994, \apj, 437.

\reference{o} Gnedin, N.Y., \& Ostriker, J.P. 1997, \apj, 486, 581.

\reference{o} Haardt, F., \& Madau, P. 1996, \apj, 461, 20.

\reference{o} Haehnelt, M.G., \& Steinmetz, M. 1997, \mnras, 289, L21.

\reference{o} Haiman, Z., \& Loeb, A. 1998a, \apj, 499, 520.

\reference{o} Haiman, Z., \& Loeb, A. 1998b, \apj, 503, 505.

\reference{o} Hellsten, U., Dave, R., Hernquist, L., Weinberg, D.H.,
\& Katz, N. 1997, \apj, 487, 482.

\reference{o} Hernquist, L., Katz, N., Weinberg, D.H.,
\& Miralda-Escud\'e, J. 1996, \apjl, 457, 51L.

\reference{o} Katz, N., Weinberg, D.H., Hernquist, L., \& 
        Miralda-Escud\'e, J. 1996, \apjl, 457, 57L.

\reference{o} Kirchhoff, H. 1860, Phil. Mag., 19, 193.

\reference{o} Madau, P., Haardt, F., \& Rees, M.J. 1998, \apj, in press
(astro-ph/9809058). 

\reference{o} Mihalas, D. \& Mihalas, B. 1984, {\em Foundations of
  Radiation Hydrodynamics}, New York: Oxford University Press.

\reference{o} Norman, M. L., Paschos, P. \& Abel, T. 1998,
in ``$H_2$ in the Early Universe", eds. F. Palla, E. Corbelli, and D. Galli, 
Memorie Della Societa Astronomica Italiana, in press (astro-ph/9807282).

\reference{o} Ostriker, J.P., \& Gnedin, N.Y. 1996, \apj, 472, L63. 

\reference{o} Paschos, P., Mihalas, D., Norman, M.L., \& Abel, T. 1998,
in preparation.

\reference{o} Razoumov, A.O., \& Scott, D. 1998, preprint (astro-ph/9810425).

\reference{o} Reimers, D., K\"ohler,S., Wisotzki, L. Groote, D.,
Rodriguez-Pascual, \& P. Wamsteker, W. 1997, \aa, 327, 890.

\reference{o} Songaila, A. 1998, \apj, 480, L1.

\reference{o} Zhang, Y., Anninos, P., \& Norman, M.L. 1995, \apj, 453, L57

\end{references}
\end{document}